\documentclass[12pt,a4paper]{article}
\usepackage{amsmath,amssymb,cite}
\usepackage[]{graphicx, epsfig}

\def\beq{\begin{equation}}
\def\eeq{\end{equation}}
\def\bea{\begin{eqnarray}}
\def\eea{\end{eqnarray}}
\def\beaa{\begin{eqnarray*}}
\def\eeaa{\end{eqnarray*}}

\def\putunder#1#2{\mathrel{
\setbox0=\hbox{#1}\setbox1=\hbox{\scriptsize #2} \dimen0=-0.5\wd0
\advance\dimen0 by -0.5\wd1 \dimen1=0.5\wd0 \advance\dimen1 by
-0.5\wd1
\hbox{\box0\kern\dimen0%
\vbox to 0pt {\hbox{\lower 0.7em \box1}\vss}%
\kern\dimen1} }}

\newcommand{\gsim}{\lower.7ex\hbox{$\;\stackrel{\textstyle>}{\sim}\;$}}
\newcommand{\lsim}{\lower.7ex\hbox{$\;\stackrel{\textstyle<}{\sim}\;$}}
\newcommand{\be}{\begin{equation}}
\newcommand{\e}{\end{equation}}

\begin{document}

\thispagestyle{empty}
\vspace*{.5cm}
\noindent
SISSA-08/2007/EP\hspace*{\fill} January 23, 2007\\
\vspace*{1.6cm}

\begin{center}
{\Large\bf Consistent de Sitter String Vacua from
\\[0.3cm]K\"ahler Stabilization and D-term uplifting}
\\[2.0cm]
{\large S. L. Parameswaran and A. Westphal}\\[.5cm]
{\it ISAS-SISSA and INFN, Via Beirut 2-4, I-34014 Trieste, Italy}
\\[3cm]

{\bf Abstract}\end{center} \noindent In this note, we review our construction of de Sitter vacua in type IIB flux compactifications, in which moduli
stabilization and D-term uplifting can be combined consistently with
the supergravity constraints. Here, the closed string fluxes fix the
dilaton and the complex structure moduli while perturbative
quantum corrections to the K\"ahler potential stabilize the volume
K\"ahler modulus in an $AdS_4$-vacuum. Then, magnetized $D7$-branes provide consistent supersymmetric
D-term uplifting towards $dS_4$. Based on hep-th/0602253.

\noindent 

{\vfill\leftline{}\vfill \vskip  10pt \footnoterule {\footnotesize
 This is the transcript of a talk given by A. W. at the RTN project `Constituents, fundamental forces and symmetries of the universe' conference in Napoli, October 9 -13, 2006.\\
 Email: param@sissa.it, westphal@sissa.it\vskip -12pt}}

\newpage

\baselineskip=10pt

\section{Introduction}

Recent developments in string theory have seen the discovery of a whole
'landscape'~\cite{BoussoP,kklt,sussk,dougl} of stable and
meta-stable 4d vacua.  This represents remarkable
progress in the formidable task of constructing realistic 4d
string vacua.  In particular, the most pressing issues have been
how to stabilize the geometrical moduli of a compactification, and
at the same time address the tiny, positive cosmological constant
that is inferred from the present-day accelerated expansion of the
universe. Recently, the use of closed string background fluxes in
string compactifications has been studied in this
context~\cite{GKP,CBachas,PolStrom,Michelson,
DasSeRa,TaylVaf,GVW,Vafa,Mayr,GSS,KlebStrass,Curio2,CKLT,HaaLou,BB,
DallAgata,KaScTr,silver,acharya,dlust}. Such flux
compactifications can stabilize the dilaton and the complex
structure moduli in type IIB string theory. Non-perturbative
effects such as the presence of $Dp$-branes~\cite{Verl} and
gaugino condensation were then used by Kachru {\it et al}
(henceforth KKLT)~\cite{kklt} to stabilize the remaining K\"ahler
moduli in such type IIB flux compactifications (for related
earlier work in heterotic M-theory see~\cite{Curio1}).
Simultaneously these vacua allow for SUSY breaking and thus the
appearance of metastable $dS_4$-minima with a small positive
cosmological constant fine-tuned in discrete steps.
KKLT~\cite{kklt} used the SUSY breaking effects of an
$\overline{D3}$-brane to achieve this. Alternatively the effect of
D-terms on $D7$-branes has been considered in this
context~\cite{bkqu}.

Bearing in mind the importance of constructing 4d de Sitter string
vacua in a reliable way, one should be aware of the problems in using
$\overline{D3}$-branes as uplifts which arise due to their explicit breaking of SUSY.  If we replace the
$\overline{D3}$-branes by D-terms driven by gauge fluxes on
$D7$-branes~\cite{bkqu} we therefore retain considerably more control as we remain inside supergravity. In this case the
requirements of both 4d supergravity and the $U(1)$ gauge
invariance necessary for the appearance of a D-term place
consistency conditions on the implementation of a D-term (noted
in~\cite{bkqu}, and emphasised in~\cite{BDKP,DuVe,VZ}). It has proven difficult to meet these conditions in a concrete stringy realisation of~\cite{bkqu}, where the proposal was made in the
context of KKLT. A consistent mechanism of stabilizing a modulus
via D-terms and uplifting its minimum to a metastable dS vacuum
has been constructed within the context of 4d supergravity
by~\cite{VZ} without, however, having a viable string embedding -
a more stringy and consistent model can be found in~\cite{DuVe}. Recent further stringy constructions were presented along the lines of~\cite{DuVe} where the model-dependent inclusion of charged matter fields renders a modified KKLT superpotential gauge invariant and provides consistent uplifting D-terms, see e.g.~\cite{Carlos,DuMamb,HaackKrefl}.

Given the subtleties encountered in the D-term uplifting of KKLT potentials, it is appealing that recently the
possibility of stabilizing the remaining K\"ahler volume modulus
of type IIB flux compactifications purely by perturbative
corrections to the K\"ahler potential has been
studied~\cite{HG,BHK2}. The leading corrections which the K\"ahler
potential receives are given by an ${\cal
O}(\alpha'^3)$-correction~\cite{bbhl} and string loop
corrections~\cite{BHK1}. The $\alpha'$-corrections have recently
been used to provide a realization of the simplest KKLT $dS$-vacua with F-term uplifting
without the need for $\overline{D3}$-branes as the source of
uplifting~\cite{Brama,Bobk,West,West2} (for other recent models using a more general O'Raifaertaigh like F-term uplifting sector see e.g.~\cite{Flift}). Under certain conditions the
interplay of both the $\alpha'$-correction and the loop
corrections leads to a stabilization of the volume modulus by the
perturbative corrections alone~\cite{BHK2}. The corrections to the
K\"ahler potential do not break the shift symmetry of the volume
modulus. Therefore, in the present note, we show that such a
K\"ahler stabilization mechanism allows for a consistent D-term
uplift, by gauging this shift symmetry with world-volume gauge
fluxes on a $D7$-brane.  Moreover, from simple scaling arguments one
can conclude that the resulting vacuum does not suffer from any
tachyonic directions.

\section{D-terms uplifts and consistency conditions from 4d
${\cal N}=1$ supergravity}\label{D-term}

The proposal to use a field dependent FI D-term as a source of
uplifting $AdS$- to $dS$-vacua was constructed in~\cite{bkqu}.
Consider a 4d ${\cal N}=1$ compactification of type IIB string
theory on an orientifolded Calabi-Yau 3-fold in the presence of
closed string fluxes. The $G_{(3)}$-flux fixes the dilaton $S$ and
the complex structure moduli $U^I$. Generically, this procedure
leaves the K\"ahler moduli unfixed and in particular the universal
K\"ahler volume modulus $T$. Now, the volume modulus enjoys a
Peccei-Quinn type symmetry: $T \rightarrow T + i\alpha$.  In the
presence of a background 2-form gauge field strength $F_{mn}$,
threading the world-volume of a $D7$-brane wrapped on a 4-cycle
$\Gamma$ of the compact internal manifold, this symmetry is
gauged.  The corresponding gauge covariant derivative acts on
$b={\rm Im}\,T$ as $D_{\mu}b=\partial_{\mu}b+iqA_{\mu}$, with $q$
the charge.  The necessary coupling, $qA^{\mu}\partial_{\mu}b$,
arises from the $a_{(2)}\wedge F_{(2)}$-coupling contained in the
world volume action of the $D7$-brane, where $b$ and $a_{(2)}$ are
dual fields. Here $a_{(2)}$ denotes the 2-form potential contained
in the closed string 4-form $C_{(4)}$ which has the world volume
coupling $C_{(4)}\wedge F_{(2)}\wedge F_{(2)}$ to the $U(1)$-gauge
field strength $F_{(2)}=dA_{(1)}$ on the $D7$-brane. Note that $q$ turns out to be quantized
due to its relation with the effective D3-brane charge carried by the $D7$-brane. As long as we
assume just one $D7$-brane its $U(1)$ world volume gauge theory
has no local anomalies.

As expected, the gauging goes hand in hand with a D-term
potential, and specifically a contribution to the scalar potential
for the volume modulus $T$.  This arises from the world-volume
action of the wrapped $D7$-brane, as: \beq V_D(T)\sim
T_7\cdot\int_{\Gamma}d^4y\sqrt{g_8}F_{mn}F^{mn} \sim
\frac{q^2}{(T+\bar{T})^3}\;\;,\label{BKQuplift} \eeq where $T_7$
is the $D7$-brane tension. For simplicity we are assuming a single
K\"ahler modulus, and also the absence of matter fields charged
under the $U(1)$ gauge group. This latter assumption may be
justified in a model with a single isolated $D7$-brane: The matter
fields arising from open strings stretching between the $D7$- and
other branes would then become very massive thus driving their
VEVs to zero. In the presence of light charged matter fields, one
must consider whether their dynamics are such as to minimise the
D-term potential at $V_D=0$, or to allow this supersymmetry
breaking contribution.\footnote{See \cite{bkqu} for more
discussion on this important point.}

Recall now that it is a shift symmetry in the K\"ahler potential $K$ of the chiral superfield $T$ \beq T\to T+ i\alpha\;:\; K=-3\ln(T+\bar{T})\;\to\; K\label{shift}\eeq that is
gauged by the $D7$-brane gauge flux. The joint
requirements of gauge invariance and local supersymmetry  place tight constraints on the possibility to have a field dependent FI D-term~\cite{BDKP,VZ}. These constraints require the function \beq
G=K+\ln|W|^2\label{G}\;,\;W:\, {\rm superpotential}\eeq  to be gauge invariant under the shift symmetry eq.~\eqref{shift}. $G$ determines
the scalar potential via \beq
V=V_F+V_D=e^G(G^{T\bar{T}}G_{T}
\overline{G_{T}}-3)+\frac{1}{2}\,({\rm
Re}\,f_{D7})^{-1}D_{T}^2\;,\; \;D_{T}=iX^{T} G_{T}=iX^T \frac{\partial G}{\partial T}.\label{Veff}\eeq $X^{T}=iq$
denotes the Killing vector of the isometry eq.~\eqref{shift} and $f_{D7}=T$ the $D7$-brane gauge kinetic function. Notice that $D_T\sim F_T = e^{G/2} G_T$, so that
a SUSY-breaking F-term is necessary to observe a non-trivial uplifting effect
from the D-term~\cite{nilles}. Then eq.~\eqref{Veff} tells us that the form \beq
W=W_{\rm flux}(S,U^I)+A\cdot e^{-a T}\label{WIIB}\;\;.\eeq that the KKLT
superpotential in type IIB flux compactifications
generically takes (for the case of
just the one universal K\"ahler modulus $T$, i.e. the volume) is gauge invariant only if either $W_{\rm flux}=0$ or $A=0$. Thus, if
background fluxes are used to stabilize $S$ and the $U^I$ and
non-perturbative effects are used to stabilize $T$, as in
KKLT~\cite{kklt}, then the shift symmetry that the K\"ahler
potential has cannot be gauged to yield a D-term
uplift~\cite{VZ}, unless further fields like hidden sector charged matter are introduced~\cite{DuVe,Carlos,DuMamb,HaackKrefl}.

Keeping the invariance under shifts
(to allow the D-term) while stabilizing $T$ demands that $T$ has
to be stabilized by corrections depending solely on $T+\bar{T}$.
By holomorphy of the superpotential we are then led to consider
stabilization of $T$ by perturbative corrections to the K\"ahler
potential, which depend only on $T+\bar{T}$.

\section{Perturbative corrections to the K\"ahler potential and
volume stabilization}\label{AdS}

Recently the possibility of stabilizing the volume modulus of type
IIB flux compactifications solely by perturbative corrections to
the K\"ahler potential has received some
attention~\cite{bbhl,HG,BHK2}. This is due to the fact that the
two leading corrections have been derived in type IIB string
theory explicitly (for a few concrete examples, at least).

Firstly, one has in type IIB compactified on an orientifolded
Calabi-Yau threefold an ${\cal O}(\alpha'^3)$ $R^4$-correction to
the 10d type IIB supergravity action~\cite{bbhl,GreenSethi} (all other corrections ${\cal
O}(\alpha'^3)$ or higher are subleading, see~\cite{Brama2,West3}).  This generates a
correction to the K\"ahler potential~\cite{bbhl}  \beq \Delta
K_{\alpha'^3}^{R^4}=-(2\pi\alpha')^3\frac{\hat{\xi}}{(T+\bar{T})^{3/2}}+{\cal
O}(\alpha'^6)\;,\;\;\;
\hat{\xi}=-\frac{1}{4\sqrt{2}}\;\zeta(3)\cdot\chi\cdot(S+\bar{S})^{3/2}
=:\xi\cdot(S+\bar{S})^{3/2} \label{dK}\;\;.\eeq Here ${\cal V}=(T+\bar{T})^{3/2}$ denotes
the Calabi-Yau volume and from now on we set $2\pi\alpha'=1$.

Next, there exist string loop corrections to the K\"ahler
potential. Ref.~\cite{HG} studied field theory loop corrections
arising in the 4d ${\cal N}=1$ supergravity after compactification
of type IIB string theory, which by dimensional analysis start
with a correction to the K\"ahler potential $\sim
(T+\bar{T})^{-2}$. The string loop corrections have been
calculated explicitly by~\cite{BHK1} for compactification of type
IIB string theory on the $T^6/\mathbb{Z}_2\times\mathbb{Z}_2$
orientifold with Hodge numbers $(h_{11},h_{21})=(3,51)$, and for
the ${\cal N}=2$ sector contribution in the $T^6/\mathbb{Z}_6'$
orientifold. In the case of non-hierarchical K\"ahler moduli it is
again the piece $\sim(T+\bar{T})^{-2}$ in $\Delta K^{(g_s)}$ which
is the relevant loop correction (for hierarchical K\"ahler moduli as e.g. in~\cite{Brama2,CQS} this can be different, see~\cite{Haack}).

Assuming that the dilaton and complex structure are stabilised,
$D_S W = 0$ and $D_U W = 0$, then the large volume expansion of
the scalar potential induced by all the above corrections starts
with~\cite{HG,BHK2} \beq
V=e^{K^{(0)}}\cdot|W|^2\cdot\left(\frac{c_1}{(T+\bar{T})^{3/2}}+
\frac{c_2}{(T+\bar{T})^2}+\ldots\right)\;,\; c_1 =
3/4\cdot\hat{\xi}\;,\; c_2 =\beta\cdot
(U+\bar{U})^2\label{VlargeT}\eeq where
the first piece is the $\alpha'$ correction, and
the second is the string loop correction with $\beta$ a constant.

Now we can see that when $c_2>0$ and $c_1<0$ (which corresponds to
$\chi>0$) and $|c_2/c_1|\gg 1$ there is inevitably a
non-supersymmetric $AdS_4$-minimum for the scalar potential of
${\rm Re}\,T$ containing both corrections at large
volume~\cite{HG} (see also~\cite{BHK2}).

Unfortunately, in the only fully calculated example,
$T^6/\mathbb{Z}_2\times\mathbb{Z}_2$, we have
$\chi=2\cdot(h_{11}-h_{21})<0$, for which there is no minimum. We
may however look to the orientifold
$T^6/\mathbb{Z}_6'$~\cite{BHK1,BHK2} as a promising candidate for
the implementation of our scenario. There, $\chi>0$ and the known
${\cal N}=2$ part of the loop corrections takes the same form as
the $T^6/\mathbb{Z}_2\times\mathbb{Z}_2$ corrections (the
inequivalent $T^6/\mathbb{Z}_2\times\mathbb{Z}_2$-orientifold also
has $\chi
>0$ but there the requirement of exotic $O$-planes~\cite{AntLoop}
may complicate the loop corrections, which are presently unknown).

Finally, we comment (following a similar discussion in~\cite{Brama2,CQS}) on the stability of the minimum found
above with respect to the minima for the complex structure moduli
and the dilaton. Prior to the introduction of the perturbative
corrections to the K\"ahler potential, the complex structure
moduli and the dilaton were fixed by background fluxes through the
conditions $D_UW=0$ and $D_SW=0$. Consider now the case that the
K\"ahler corrections, which are negative near to the minimum of
$T$, try to drive away $S$ and/or $U$ from their minima
$D_SW=D_UW=0$. Then the tree level flux potential yields a
contribution $V_{\rm flux}\sim {\cal O}(+\frac{1}{{\cal V}^2})$,
while the K\"ahler corrections contribute at ${\cal
O}(-\frac{1}{{\cal V}^3})$, which is subleading at large volumes.
Thus, the corrections cannot destabilize the original minima of
$S$ and $U$ and these remain minima of the full theory including
the K\"ahler corrections which stabilize $T$. This is an improvement on KKLT where stability may fail, see e.g.~\cite{nilles2}. 
Moreover, similar arguments can of course be used after including
the uplifting, to which we now turn.

\section{de Sitter vacua from a consistent D-term}\label{dS}

It is now easy to see that the perturbative $AdS_4$-minimum for
$T$ discussed in the last Section can be uplifted to a
$dS_4$-minimum with a consistent D-term. The $AdS_4$-minimum is non-supersymmetric.  Moreover, the full theory including
the perturbative corrections to the K\"ahler potential is a
function of $T+\bar{T}$ alone. Thus it is fully invariant under
the shift $T\to T+i\alpha$ and in particular we have invariance of
$G=K+\ln|W|^2$ under this shift symmetry. Therefore, the mechanism
of K\"ahler stabilization of the volume modulus $T$ fulfills the
consistency constraints of Sect.~\ref{D-term}. This allows us to
gauge the shift symmetry, using world-volume fluxes on a
$D7$-brane, as described in that section.

The full scalar potential will now contain a D-term piece in
addition to the F-term contributions from the K\"ahler
corrections. For the $T^6/\mathbb{Z}_6'$ example the potential,
expanded up to ${\cal O}(\alpha'^3/(T+\bar{T})^{3/2})$ and to
leading order in the string loop corrections, reads \bea
V=V_F+V_D&=& \frac{|W|^2}{(T+\bar{T})^3}\cdot\left(\frac{c_1}{(T+\bar{T})^{3/2}}
+\frac{c_2}{(T+\bar{T})^2}\right)+\frac{9\,q^2}{(T+\bar{T})^3}\label{VFD}\eea where the constants $c_1$ and $c_2$ are
evaluated in the minima of $U$ and $S$
determined by $D_UW=D_SW=0$. Note that $V_D$ has been expanded only to leading
order since later tuning will require $V_D$ to cancel $V_F$ to
leading order. Taking into account the higher orders in $D_T$
would require to write the higher orders in $V_F$ as well for
consistency.\footnote{For further details see~\cite{West3}.}

\begin{figure}[ht]
\begin{center}
\includegraphics[width=12cm]{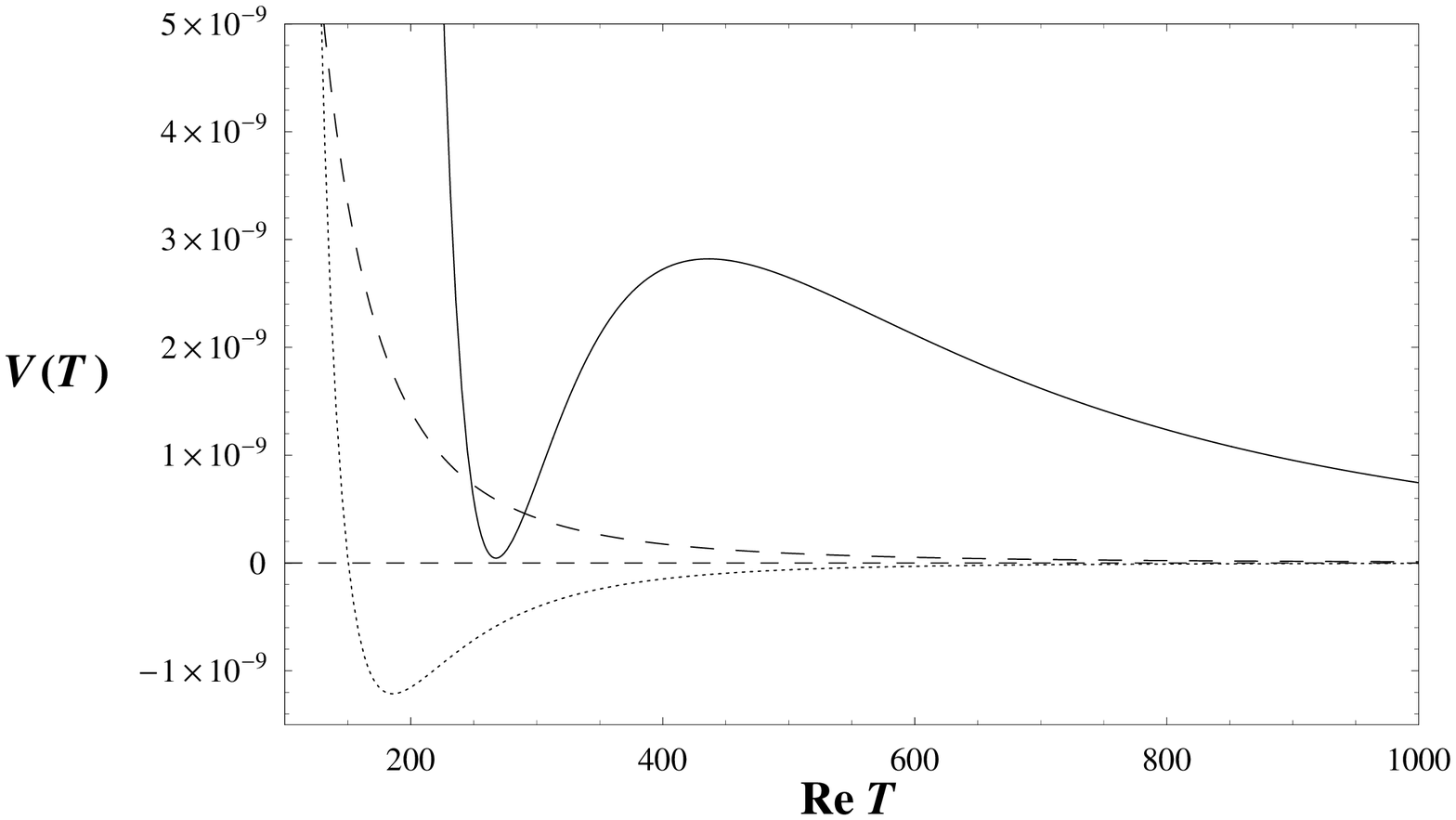}
\end{center}
\refstepcounter{figure}\label{Fig.1}

\vspace*{-.2cm} {\bf Figure~\ref{Fig.1}:} Dotted: The F-term
scalar potential $V_F(T)$ leading to perturbative K\"ahler
stabilization of $T$. Dashed: The uplifting D-term scalar
potential $V_D(T)$. Both graphs have been rescaled by $10^{-2}$
for display reasons. Solid: The scalar potential eq.~\eqref{VFD}
after uplifting by switching on a gauge field background on a
single $D7$-brane. The numbers are chosen in this example as
$W_0=25.5$, $q=1$, ${\rm Re}\,U=242$, ${\rm Re}\,S=10$ and
$\chi=48$. Also $\beta=1/(2\pi^2)$ is taken from the
$T^6/\mathbb{Z}_2\times\mathbb{Z}_2$ as a guiding example.

\end{figure}

Given that ${\rm Re}\,T$ is stabilized at large volume ${\cal V} =
(T + \bar{T})^{3/2}$, and assuming that $q^2$ is ${\cal O}(1)$, we
can arrange for a situation where \beq \Big|\left.V_F\right|_{\rm
min}\Big|\sim \frac{|W|^2}{{\cal V}^3}\sim \left. V_D\right|_{\rm
min}\sim\frac{q^2}{{\cal V}^2}\label{tuning}\eeq holds by tuning
$W$ to larger values such that we get $V_F+V_D\approx 0$ in the
minimum.

This situation is displayed in Fig.~\ref{Fig.1} using the
potential given in eq.~\eqref{VFD} for the semi-explicit
$T^6/\mathbb{Z}_6'$-example. This serves as an indication of how
we expect the behavior to be in a fully explicit model. In any
case, given the vast landscape of type IIB flux compactifications,
we expect that there should be many models in type IIB string
theory where our uplifting scenario yields qualitatively the same
results as discussed here. We finally mention here that in our scenario the gravitino mass is typically large, i.e. $m_{3/2}=e^{K/2}|W|\sim10^{-2}\ldots 10^{-3}\,M_P$ since $W\gsim {\cal O}(1)$ and ${\cal V}\lsim {\cal O}(10^3)$ here (see e.g.~\cite{ChoiJeong} for similar results in KKLT based models).

\section{Conclusion}\label{con}

In this note we discussed a mechanism for generating de Sitter
vacua in string theory by spontaneously breaking supersymmetry
with consistent D-terms. This proposal has proven difficult to
consistently embed in a stringy scenario.\footnote{Note added: for 'hot off
the plate' progress on explicit constructions in other
scenarios see~\cite{Qbkq2}.}  We find that type IIB
flux compactifications, with volume stabilization via perturbative
corrections to the K\"ahler potential, provide such a scenario. As
discussed in the literature, $\alpha'$- and string loop
corrections allow for stabilization of the $T$ modulus by purely
perturbative means without turning to non-perturbative effects
such as gaugino condensation. The
K\"ahler corrections preserve the invariance of the theory under a
shift symmetry of the $T$ modulus. In the presence of a magnetised
$D7$-brane this unbroken shift invariance is gauged which leads to
supersymmetric D-terms from string theory which fulfill all the
known consistency requirements of 4d ${\cal N}=1$ supergravity.
These D-terms then provide a parametrically small and tunable
uplift of the perturbatively stabilized $AdS_4$-minimum towards a
metastable $dS$-minimum. In view of the desire to search the
'landscape' of string theory vacua for those regions where
spontaneously broken supersymmetry allows for certain control of
the low-energy effective theory, the discussed mechanism of a
consistent D-term uplift in string theory promises access to a new
class of metastable $dS$-vacua. The SUSY breaking provided in this scenario is typically a high-scale one with $m_{3/2}\sim10^{-2}\ldots 10^{-3}\,M_P$.  In a fully explicit model, it
would be necessary to calculate the string loop corrections in the
presence of gauged symmetries and magnetised D-branes, for example
along the lines of \cite{gaugedloops}.  It would also be
interesting to study the consequences of this uplifting mechanism
for possible realizations of inflation in string theory, as well
as the low-energy phenomenology of this type of spontaneous SUSY
breaking.

\vspace*{1cm} \noindent {\bf Acknowledgements}:
We would like to thank Z. Tavartkiladze for useful
discussions and comments.


\begin{thebibliography}{99}
\bibitem{BoussoP}
R.~Bousso \& J.~Polchinski, JHEP {\bf 0006}, 006 (2000)
[arXiv:hep-th/0004134].
%%CITATION = HEP-TH 0004134;%%

\bibitem{kklt}
S.~Kachru, R.~Kallosh, A.~Linde \& S.~P.~Trivedi, Phys.\ Rev.\ D
{\bf 68}, 046005 (2003) [arXiv:hep-th/0301240].
%%CITATION = HEP-TH 0301240;%%

\bibitem{sussk}
L.~Susskind, [arXiv:hep-th/0302219].
%%CITATION = HEP-TH 0302219;%%

\bibitem{dougl}
M.~R.~Douglas, JHEP {\bf 0305}, 046 (2003)
[arXiv:hep-th/0303194].
%%CITATION = HEP-TH 0303194;%%

\bibitem{GKP}
S.~B.~Giddings, S.~Kachru \& J.~Polchinski, Phys.\ Rev.\ D {\bf
66}, 106006 (2002) [arXiv:hep-th/0105097].
%%CITATION = HEP-TH 0105097;%%

\bibitem{CBachas}
C.~Bachas, [arXiv:hep-th/9503030].
%%CITATION = HEP-TH 9503030;%%

\bibitem{PolStrom}
J.~Polchinski \& A.~Strominger, Phys.\ Lett.\ B {\bf 388} (1996)
736 [arXiv:hep-th/9510227].
%%CITATION = HEP-TH 9510227;%%

\bibitem{Michelson}
J.~Michelson, Nucl.\ Phys.\ B {\bf 495} (1997) 127
[arXiv:hep-th/9610151].
%%CITATION = HEP-TH 9610151;%%

\bibitem{DasSeRa}
K.~Dasgupta, G.~Rajesh \& S.~Sethi, JHEP {\bf 9908} (1999) 023
[arXiv:hep-th/9908088].
%%CITATION = HEP-TH 9908088;%%

\bibitem{TaylVaf}
T.~R.~Taylor \& C.~Vafa, Phys.\ Lett.\ B {\bf 474} (2000) 130
[arXiv:hep-th/9912152].
%%CITATION = HEP-TH 9912152;%%

\bibitem{GVW}
S.~Gukov, C.~Vafa \& E.~Witten, Nucl.\ Phys.\ B {\bf 584}, 69
(2000) [Erratum-ibid.\ B {\bf 608}, 477 (2001)]
[arXiv:hep-th/9906070].
%%CITATION = HEP-TH 9906070;%%

\bibitem{Vafa}
C.~Vafa, J.\ Math.\ Phys.\  {\bf 42}, 2798 (2001)
[arXiv:hep-th/0008142].
%%CITATION = HEP-TH 0008142;%%

\bibitem{Mayr}
P.~Mayr, Nucl.\ Phys.\ B {\bf 593} (2001) 99
[arXiv:hep-th/0003198].
%%CITATION = HEP-TH 0003198;%%

\bibitem{GSS}
B.~R.~Greene, K.~Schalm \& G.~Shiu, Nucl.\ Phys.\ B {\bf 584}
(2000) 480 [arXiv:hep-th/0004103].
%%CITATION = HEP-TH 0004103;%%

\bibitem{KlebStrass}
I.~R.~Klebanov \& M.~J.~Strassler, JHEP {\bf 0008}, 052 (2000)
[arXiv:hep-th/0007191].
%%CITATION = HEP-TH 0007191;%%

\bibitem{Curio2}
G.~Curio \& A.~Krause, Nucl.\ Phys.\ B {\bf 602}, 172 (2001)
[arXiv:hep-th/0012152].
%%CITATION = HEP-TH 0012152;%%

\bibitem{CKLT}
G.~Curio, A.~Klemm, D.~L{\"u}st \& S.~Theisen, Nucl.\ Phys.\ B
{\bf 609} (2001) 3 [arXiv:hep-th/0012213];\\
%%CITATION = HEP-TH 0012213;%%
G.~Curio, A.~Klemm, B.~K{\"o}rs \& D.~L{\"u}st, Nucl.\ Phys.\ B
{\bf 620} (2002) 237 [arXiv:hep-th/0106155].
%%CITATION = HEP-TH 0106155;%%

\bibitem{HaaLou}
M.~Haack \& J.~Louis, Nucl.\ Phys.\ B {\bf 575} (2000) 107
[arXiv:hep-th/9912181]; \\
%%CITATION = HEP-TH 9912181;%%
Phys.\ Lett.\ B {\bf 507} (2001) 296 [arXiv:hep-th/0103068].
%%CITATION = HEP-TH 0103068;%%

\bibitem{BB}
K.~Becker \& M.~Becker, JHEP {\bf 0107} (2001) 038
[arXiv:hep-th/0107044].
%%CITATION = HEP-TH 0107044;%%

\bibitem{DallAgata}
G.~Dall'Agata, JHEP {\bf 0111} (2001) 005 [arXiv:hep-th/0107264].
%%CITATION = HEP-TH 0107264;%%

\bibitem{KaScTr}
S.~Kachru, M.~B.~Schulz \& S.~Trivedi, JHEP {\bf 0310} (2003) 007
[arXiv:hep-th/0201028].
%%CITATION = HEP-TH 0201028;%%

\bibitem{silver}
E.~Silverstein, [arXiv:hep-th/0106209];\\
%%CITATION = HEP-TH 0106209;%%
A.~Maloney, E.~Silverstein \& A.~Strominger,
[arXiv:hep-th/0205316].
%%CITATION = HEP-TH 0205316;%%

\bibitem{acharya}
B.~S.~Acharya, [arXiv:hep-th/0212294].
%%CITATION = HEP-TH 0212294;%%

\bibitem{dlust}
R.~Blumenhagen, D.~L\"ust \& T.~R.~Taylor, Nucl.\ Phys.\ B {\bf
663}, 319 (2003)
[arXiv:hep-th/0303016].\\
%%CITATION = HEP-TH 0303016;%%
J.~F.~G.~Cascales \& A.~M.~Uranga, JHEP {\bf 0305}, 011 (2003)
[arXiv:hep-th/0303024].
%%CITATION = HEP-TH 0303024;%%

\bibitem{Verl}
H.~Verlinde, Nucl.\ Phys.\ B {\bf 580}, 264 (2000)
[arXiv:hep-th/9906182].
%%CITATION = HEP-TH 9906182;%%

\bibitem{Curio1}
G.~Curio \& A.~Krause, Nucl.\ Phys.\ B {\bf 643}, 131 (2002)
[arXiv:hep-th/0108220].
%%CITATION = HEP-TH 0108220;%%

\bibitem{bkqu}
C.~P.~Burgess, R.~Kallosh \& F.~Quevedo, JHEP {\bf 0310}, 056
(2003) [arXiv:hep-th/0309187].
%%CITATION = HEP-TH 0309187;%%

\bibitem{BDKP}
P.~Binetruy, G.~Dvali, R.~Kallosh \& A.~Van Proeyen, Class.\
Quant.\ Grav.\  {\bf 21}, 3137 (2004) [arXiv:hep-th/0402046].
%%CITATION = HEP-TH 0402046;%%

\bibitem{DuVe}
E.~Dudas \& S.~K.~Vempati, Nucl.\ Phys.\ B {\bf 727}, 139 (2005)
[arXiv:hep-th/0506172].
%%CITATION = HEP-TH 0506172;%%

\bibitem{VZ}
G.~Villadoro \& F.~Zwirner, Phys.\ Rev.\ Lett.\  {\bf 95}, 231602
(2005) [arXiv:hep-th/0508167].
%%CITATION = HEP-TH 0508167;%%

\bibitem{Carlos}
A.~Achucarro, B.~de Carlos, J.~A.~Casas \& L.~Doplicher,
[arXiv:hep-th/0601190].
%%CITATION = HEP-TH 0601190;%%

\bibitem{DuMamb}
E.~Dudas \& Y.~Mambrini, JHEP {\bf 0610}, 044 (2006)
[arXiv:hep-th/0607077].
%%CITATION = HEP-TH 0607077;%%

\bibitem{HaackKrefl}
M.~Haack, D.~Krefl, D.~Lust, A.~Van Proeyen \& M.~Zagermann, [arXiv:hep-th/0609211].
%%CITATION = HEP-TH 0609211;%%

\bibitem{HG}
G.~von Gersdorff \& A.~Hebecker, Phys.\ Lett.\ B {\bf 624}, 270
(2005) [arXiv:hep-th/0507131].
%%CITATION = HEP-TH 0507131;%%

\bibitem{BHK2}
M.~Berg, M.~Haack \& B.~Kors, Phys.\ Rev.\ Lett.\  {\bf 96},
021601 (2006) [arXiv:hep-th/0508171].
%%CITATION = HEP-TH 0508171;%%

\bibitem{bbhl}
K.~Becker, M.~Becker, M.~Haack \& J.~Louis, JHEP {\bf 0206}, 060
(2002) [arXiv:hep-th/0204254].
%%CITATION = HEP-TH 0204254;%%

\bibitem{BHK1}
M.~Berg, M.~Haack \& B.~Kors, JHEP {\bf 0511}, 030 (2005)
[arXiv:hep-th/0508043].
%%CITATION = HEP-TH 0508043;%%

\bibitem{Brama}
V.~Balasubramanian \& P.~Berglund, JHEP {\bf 0411}, 085 (2004)
[arXiv:hep-th/0408054].
%%CITATION = HEP-TH 0408054;%%

\bibitem{Bobk}
K.~Bobkov, JHEP {\bf 0505}, 010 (2005) [arXiv:hep-th/0412239].
%%CITATION = HEP-TH 0412239;%%

\bibitem{West}
A.~Westphal, JCAP {\bf 0511}, 003 (2005) [arXiv:hep-th/0507079].
%%CITATION = HEP-TH 0507079;%%

\bibitem{West2}
A.~Westphal, 2006 [arXiv:hep-th/0611332].
%%CITATION = HEP-TH 0611332;%%

\bibitem{Flift}
K.~Intriligator, N.~Seiberg \& D.~Shih, JHEP {\bf 0604}, 021
(2006) [arXiv:hep-th/0602239];\\{}
%%CITATION = HEP-TH 0602239;%%
M.~Gomez-Reino \& C.~A.~Scrucca, JHEP {\bf 0605}, 015 (2006)
[arXiv:hep-th/0602246];\\{}
%%CITATION = HEP-TH 0602246;%%
O.~Lebedev, H.~P.~Nilles \& M.~Ratz, Phys.\ Lett.\ B {\bf 636}, 126 (2006) [arXiv:hep-th/0603047];\\{}
%%CITATION = HEP-TH 0603047;%%
F.~Br\"ummer, A.~Hebecker \& M.~Trapletti, Nucl.\ Phys.\ B {\bf
755}, 186 (2006) [arXiv:hep-th/0605232];\\{}
%%CITATION = HEP-TH 0605232;%%
E.~Dudas, C.~Papineau \& S.~Pokorski, [arXiv:hep-th/0610297];\\{}
%%CITATION = HEP-TH 0610297;%%
H.~Abe, T.~Higaki, T.~Kobayashi \& Y.~Omura,
[arXiv:hep-th/0611024];\\{}
%%CITATION = HEP-TH 0611024;%%
R.~Kallosh \& A.~Linde, [arXiv:hep-th/0611183].
%%CITATION = HEP-TH 0611183;%%

\bibitem{nilles}
K.~Choi, A.~Falkowski, H.~P.~Nilles \& M.~Olechowski, Nucl.\ Phys.\ B {\bf 718}, 113 (2005) [arXiv:hep-th/0503216].
%%CITATION = HEP-TH 0503216;%%

\bibitem{GreenSethi}
M.~B.~Green \& S.~Sethi, Phys.\ Rev.\ D {\bf 59}, 046006 (1999)
[arXiv:hep-th/9808061].
%%CITATION = HEP-TH 9808061;%%

\bibitem{Haack}
M.~Haack, talk at KITP, Santa Barbara, August 15, 2006.

\bibitem{AntLoop}
C.~Angelantonj, I.~Antoniadis, G.~D'Appollonio, E.~Dudas \&
A.~Sagnotti, Nucl.\ Phys.\ B {\bf 572}, 36 (2000)
[arXiv:hep-th/9911081].
%%CITATION = HEP-TH 9911081;%%

\bibitem{Brama2}
V.~Balasubramanian, P.~Berglund, J.~P.~Conlon \& F.~Quevedo, JHEP
{\bf 0503}, 007 (2005) [arXiv:hep-th/0502058].
%%CITATION = HEP-TH 0502058;%%

\bibitem{CQS}
J.~P.~Conlon, F.~Quevedo \& K.~Suruliz, JHEP {\bf 0508}, 007
(2005) [arXiv:hep-th/0505076].
%%CITATION = HEP-TH 0505076;%%

\bibitem{nilles2}
K.~Choi, A.~Falkowski, H.~P.~Nilles, M.~Olechowski \&
S.~Pokorski, JHEP {\bf 0411} (2004) 076 [arXiv:hep-th/0411066];\\{}
%%CITATION = HEP-TH 0411066;%%
S.~P.~de Alwis, Phys.\ Lett.\ B {\bf 626}, 223 (2005) [arXiv:hep-th/0506266].
%%CITATION = HEP-TH 0506266;%%

\bibitem{West3}
S.~L.~Parameswaran \& A.~Westphal, JHEP {\bf 0610}, 079 (2006) [arXiv:hep-th/0602253].
%%CITATION = HEP-TH 0602253;%%

\bibitem{ChoiJeong}
K.~Choi \& K.~S.~Jeong, JHEP {\bf 0608}, 007 (2006)
[arXiv:hep-th/0605108].
%%CITATION = HEP-TH 0605108;%%

\bibitem{Qbkq2}
D.~Cremades, M.~P.~G.~del Moral, F.~Quevedo \& K.~Suruliz, [arXiv:hep-th/0701154].
%%CITATION = HEP-TH 0701154;%%

\bibitem{gaugedloops}
D.~Lust \& S.~Stieberger, [arXiv:hep-th/0302221].
%%CITATION = HEP-TH 0302221;%%



\end{thebibliography}
\end{document}